\documentclass[conference]{IEEEtran}
\IEEEoverridecommandlockouts

\usepackage{cite}
\usepackage{amsmath,amssymb,amsfonts}
\usepackage{algorithmic}
\usepackage{graphicx}
\usepackage{textcomp}
\usepackage{xcolor}
\def\BibTeX{{\rm B\kern-.05em{\sc i\kern-.025em b}\kern-.08em
    T\kern-.1667em\lower.7ex\hbox{E}\kern-.125emX}}

\usepackage{algorithm}
\usepackage{booktabs}

\usepackage{array}

\usepackage{float}
\begin{document}

\title{Statistical Verification of Medium-Access Parameterization for Power-Grid Edge Ad Hoc Sensor Networks}

\author{
\IEEEauthorblockN{
Haitian Wang\textsuperscript{1},
Xinyu Wang\textsuperscript{1},
Zichen Geng\textsuperscript{1},
Xian Zhang\textsuperscript{1},
Yiren Wang\textsuperscript{1},
Yihao Ding\textsuperscript{1}\IEEEauthorrefmark{1}\thanks{\IEEEauthorrefmark{1}Corresponding author: Yihao Ding (yihao.ding@uwa.edu.au).}
}
\IEEEauthorblockA{
\textsuperscript{1}Department of Computer Science and Software Engineering, School of Physics, Mathematics and Computing,\\
The University of Western Australia, 35 Stirling Highway, Crawley, WA 6009, Australia\\
E-mails: haitian.wang@uwa.edu.au, xinyu.wang@uwa.edu.au, zen.geng@research.uwa.edu.au,\\
xian.zhang@research.uwa.edu.au, bohr.wang@uwa.edu.au, yihao.ding@uwa.edu.au
}
}

\maketitle
\begin{abstract}
The widespread deployment of power grid ad hoc sensor networks based on IEEE~802.15.4 raises reliability challenges when nodes selfishly adapt CSMA/CA parameters to maximize individual performance. Such behavior degrades reliability, energy efficiency, and compliance with strict grid constraints. Existing analytical and simulation approaches often fail to rigorously evaluate configurations under asynchronous, event-driven, and resource-limited conditions. We develop a verification framework that integrates stochastic timed hybrid automata with statistical model checking (SMC) with confidence bounds to formally assess CSMA/CA parameterizations under grid workloads. By encoding node- and system-level objectives in temporal logic and automating protocol screening via large-scale statistical evaluation, the method certifies Nash equilibrium strategies that remain robust to unilateral deviations. In a substation-scale scenario, the certified equilibrium improves utility from 0.862 to 0.914 and raises the delivery ratio from 89.5\% to 93.2\% when compared with an aggressive tuning baseline. Against a delivery-oriented baseline, it reduces mean per-cycle energy from 152.8~mJ to 149.2~mJ while maintaining comparable delivery performance. Certified configurations satisfy latency, reliability, and energy constraints with robustness coefficients above 0.97 and utility above 0.91.
\end{abstract}


\vspace{-2mm}
\section{Introduction}
\vspace{-2mm}

The rapid development of smart grids has driven large-scale deployment of IoT sensor networks across all segments of electric power systems \cite{mehmood2021edge, kojonsaari2023development, sarwat2017trends, abir2021iot}, enabling real-time monitoring and control through dense installations of intelligent edge devices such as wireless fault indicators, transformer monitors, etc. \cite{kirmani2022survey, bagdadee2020iot, goudarzi2022survey}. In practice, grid operators, such as the State Grid Corporation of China \cite{yang2024construction}, deploy multi-hop ad-hoc wireless sensor networks based on the IEEE 802.15.4 standard \cite{ieee8021542020,zheng2004ieee802154,gungor2006electricautomation,wang2024multispectral,wang2025p2mfds}, with communication coordinated by the CSMA/CA protocol \cite{zhao2021development,pollin2008slotted802154,park2010adaptive,wang2026bawseg}. While this architecture achieves decentralized operation and flexible connectivity, it also introduces new challenges: each node autonomously selects key protocol parameters, such as backoff, retransmission settings, etc. \cite{barbosa2025multi}, which can incentivize selfish adaptation to optimize local objectives \cite{nagai2024improve}. Such behavior undermines sensor network reliability, energy efficiency, and fairness, especially in dynamic, event-driven power grid environments \cite{sadek2021identifying,lei2024reinforcement,kyasanur2005selfishmac,ciceklidag2024high}. As a result, there is a critical need for systematic methods that can rigorously evaluate, screen, and certify protocol parameterizations \cite{alsafran2025challenges}, ensuring robust and reliable network operation even in the presence of selfish strategies and strict grid-mandated constraints \cite{khan2021reliable}.

The challenge of selfish behavior in distributed sensor networks has been examined in areas such as medium access and power control \cite{kaddi2024energy, sammour2024intelligent, yan2011selfish,ibrahim2025multistream}. Game-theoretic approaches, including Nash equilibrium analysis \cite{dai2024survey, zhang2025pbn}, provide insight into strategic adaptation but usually rely on restrictive assumptions such as synchrony, homogeneous nodes, or static topologies \cite{garnaev2024aloha}. These assumptions make them unsuitable for the asynchronous, burst-driven, and energy-constrained conditions of grid IoT deployments \cite{mohamed2022multi, bataihah2023improving}. Simulation-based studies relax some of these constraints but often lack statistical rigor and formal guarantees, limiting their use for certifying protocol robustness in operational environments \cite{ming2023improved, singh2021adaptive}. 


To address these requirements, we develop a verification pipeline for multi-hop IEEE~802.15.4 CSMA/CA parameterization in power-grid edge sensor networks. The core is a stochastic timed hybrid automata model that captures asynchronous contention, probabilistic retransmissions, and per-state energy accounting under grid workloads, including periodic reporting and event-triggered alarms. Over this model, we formalize node- and network-level objectives using cost-bounded temporal properties, and estimate the resulting utilities via statistical model checking. We then perform automated screening over an admissible parameter set and statistically certify protocol settings that remain stable against unilateral deviations, yielding a grid-constrained $\Delta$-relaxed equilibrium notion aligned with operational risk tolerance. The evaluation shows that the certified configurations satisfy latency, reliability, and energy constraints with tight uncertainty bounds, while maintaining robustness coefficients above $0.97$ and utility above $0.91$ under the studied substation-scale scenario.

\section{Related Work}

Research on selfish behavior and access control in distributed wireless networks spans analytical game-theoretic models and simulation-based evaluations, but both lines fall short for power IoT deployments. Game-theoretic analyses of medium access (often via NE) typically impose synchronized operation, static topologies, and discrete/slotted channel access \cite{sammour2024intelligent, sadek2022modeling, boujnoui2022stochastic, tarzjani2025computing, saibharath2023selfish}, which cannot capture the asynchrony, bursty traffic, and stochastic energy constraints of real grid sensor networks \cite{sadkhan2021game}. Moreover, IEEE~802.15.4 CSMA/CA introduces asynchronous backoff, probabilistic retransmissions, and heterogeneous parameter settings across devices, yielding interactions that are difficult to express in tractable closed forms, especially under dynamic duty cycles and event-triggered loads \cite{sun2021applications, chen2021deep}. Simulation studies offer flexibility \cite{saibharath2023selfish, van2021bayesian}, yet commonly lack formalized operational objectives and statistical guarantees, and they scale poorly when exploring high-dimensional parameter spaces \cite{ismail2024detecting, li2024gcn}. Standard Monte Carlo reporting of averages provides limited assurance about stability under time-bounded or probabilistic requirements \cite{tong2021throughput, tarzjani2025computing}, and typical simulators seldom integrate domain constraints such as per-node energy budgets, grid-driven schedules, or mandated availability targets \cite{sarkar2021mobile, al2024integrating}. 

\vspace{-2mm}
\section{Preliminaries and Modeling Approach}

This section outlines the formal modeling framework that underpin our protocol analysis. We introduce the STHA as a domain-adapted formalism for capturing the operational dynamics of power grid IoT sensor networks, incorporating both protocol-specific and grid-specific constraints. We then present the compositional system model, formalize the key objectives using CBTL, and articulate the game-theoretic formulation.

\subsection{Stochastic Timed Hybrid Automata}

To rigorously model the protocol execution, energy consumption, and event-driven behavior of edge sensor networks in smart grids, we propose the STHA framework extended from WTA \cite{arcile2022timed,alur1994timedautomata}. This domain-specific extension is tailored to capture critical operational constraints of field-deployed power system sensors. Each sensor node is represented by an automaton:
\begin{equation}
\mathcal{S} = (L, \ell_0, X, E, R, I, C)
\label{eq:stha}
\end{equation}
where $L$ denotes the set of discrete locations (e.g., Idle, CCA, Backoff, Transmit, Sleep), $\ell_0$ is the initial state, $X$ is a set of real-valued clocks (such as $x_{bo}$ for backoff, $x_{cca}$ for carrier sensing, $x_{tx}$ for transmission duration), $E$ is the set of transitions with guards and resets, $R$ is the rate vector for clock evolution, $I$ is the set of location invariants, and $C$ is a vector of cost variables including cumulative energy consumption ($E_{tot}$), duty cycle ($\eta$), and retransmission count ($N_{rt}$).

Our STHA model incorporates power grid–specific features that are absent from generic automata. First, it accounts for variable supply voltages and per-mode power draw $(P_{tx}, P_{rx}, P_{idle}, P_{sleep})$, enabling accurate energy accounting under deployment conditions. Second, it includes explicit duty-cycle management with event-driven preemption for grid alarms and regulated reporting intervals (e.g., $T_{int}=45$~s or grid-adaptive). Third, it models event transitions for grid faults or external triggers that enforce system-wide urgent reporting and blackout-recovery behavior, consistent with grid reliability requirements. The state of each node is $(l, v, e)$, where $l \in L$, $v$ is the clock valuation, and $e$ encodes cost variable status at runtime.

\vspace{-2mm}
\subsection{Compositional Network Model and Medium}

The overall system is formalized as a parallel composition of $N$ node STHAs and a shared wireless medium automaton:
\begin{equation}
\mathcal{G}(\vec{\theta}) = \mathcal{S}_1(\theta_1) \parallel \cdots \parallel \mathcal{S}_N(\theta_N) \parallel \mathcal{M}
\end{equation}
where $\mathcal{S}_i(\theta_i)$ is the STHA for node $i$ under strategy configuration $\theta_i$ (e.g., backoff exponents, CCA parameters, maximum retransmissions, event response policy), and $\mathcal{M}$ models contention, physical interference, and regulatory constraints (e.g., blackout recovery $\gamma_{rec}$, sensor availability $\rho_{min}$).

Transitions and rates may be stochastic, reflecting real-world heterogeneity, clock drift, and bursty grid-induced events. The medium automaton synchronizes event broadcasts, maintains channel state, and coordinates node interactions during grid-triggered emergency periods.

\subsection{Cost-Bounded Temporal Logic for System Objectives}

Performance, correctness, and operational requirements are formally expressed using an extended CBTL, generalizing standard PWCTL to our scenario \cite{kwiatkowska2022probabilistic}. For an individual node $i$, a typical specification is:
\begin{equation}
\varphi_i: \quad \Diamond_{E_{tot,i} \leq E_{lim}} \left( x_{tx,i} \leq T_{max} \wedge S_{succ,i} \geq 1 \right)
\end{equation}
stating that node $i$ must successfully deliver at least one report within $T_{max}$ and not exceed cumulative energy $E_{lim}$ in a reporting cycle. For coordinated network-level event response:
\begin{equation}
\psi: \quad \Box \left( \text{Event}_{grid} \rightarrow \Diamond_{\leq T_{evt}} \left( \sum_{i=1}^N S_{succ,i} \geq N_{req} \right) \right)
\end{equation}
meaning that for any grid event, at least $N_{req}$ nodes report successfully within or equal to $T_{evt}$.

\subsection{Formal Problem Statement: Game and Nash Equilibrium under SCMC}

We frame the protocol adaptation and selfishness mitigation as a symmetric game over $N$ nodes, each choosing a strategy $\theta_i$ from a finite parameter set $\mathcal{P}$. The utility function for node $i$ is:
\begin{equation}
U_i(\theta_1, \ldots, \theta_N) = \Pr\left[ \mathcal{G}(\theta_1, \ldots, \theta_N) \models \varphi_i \right]
\end{equation}

A strategy profile $\theta^* \in P$ is defined as a Nash Equilibrium (NE) if, for every node $i$ and any alternative $\theta'_i$, the utility satisfies $U_i(\theta^*, \ldots, \theta^*) \geq U_i(\theta'_i, \theta^*_{-i})$, where $\theta^*_{-i}$ denotes the strategies of all nodes except $i$.

In practice, we certify an approximate equilibrium notion that tolerates residual stochastic uncertainty and modeling variability. A symmetric profile $\theta^\star \in \mathcal{P}$ is a $\Delta$-relaxed Nash equilibrium if, for every node $i$ and any deviation $\theta_i' \in \mathcal{P}$,
\begin{equation}
U_i(\theta^\star,\ldots,\theta^\star) \ge \Delta \cdot U_i(\theta_i', \theta^\star_{-i}),
\end{equation}
where $\Delta \in (0,1]$ quantifies robustness against unilateral deviations. Values of $\Delta$ closer to 1 indicate stronger stability.

Due to scalability limits of exact model checking, we utilize the SCMC method to estimate utility values across $M$ randomized simulation runs:
\vspace{-2mm}
\begin{equation}
\hat{U}_i(\vec{\theta})=\frac{1}{M}\sum_{j=1}^{M}\mathbb{I}\!\left\{\text{run}_j \models \varphi_i\right\},
\end{equation}
\vspace{-1mm}
where $\mathbb{I}\{\cdot\}$ is the indicator function (1 if the $j$-th execution satisfies $\varphi_i$, 0 otherwise). This Monte Carlo estimator is unbiased under independent runs, with asymptotic variance $\hat{U}_i(1-\hat{U}_i)/M$, which we use to report $(1-\alpha)$ confidence intervals in the certification stage.

\vspace{-2mm}
\section{Statistical Screening and Certification of Nash Equilibrium Protocol Configurations}

To statistically certify protocol configurations robust to selfish behavior in power grid IoT sensor networks, we adopt a two-stage framework. The method integrates STHA modeling with SCMC to assess whether IEEE~802.15.4 CSMA/CA parameterizations satisfy $\Delta$-relaxed NE under grid-specific constraints, including energy budgets, blackout recovery, and alarm responsiveness, ensuring certified configurations remain practical and resilient.

\subsection{Stage 1: Utility Matrix Construction \& Protocol Screening}

In this stage, we do not enumerate all Nash equilibria, but instead focus on performance-driven screening of protocol candidates based on their statistical utility and compliance with grid-side objectives. Let $\mathcal{P}$ denote the admissible set of node protocol configurations, reflecting allowable CSMA/CA parameters (e.g., backoff windows $[BE_{min}, BE_{max}]$, CCA timing, maximum retry counts $N_{retry}^{max}$, duty-cycle schedule $\Omega_{duty}$, and grid event response modes). For each candidate pair $(\vartheta', \vartheta) \in \mathcal{P} \times \mathcal{P}$, the empirical utility for node $k$ is defined as the proportion of STHA-SCMC executions in which node $k$, operating under protocol $\vartheta'$ against $(N-1)$ nodes using $\vartheta$, satisfies the grid operational predicate $\phi_k$. In our setting, $\phi_k$ encodes timely delivery for periodic status reporting and event-triggered alarms, compliance with a per-cycle energy budget, and an availability requirement during blackout-recovery intervals.

The screening procedure incrementally evaluates protocol pairs and prunes dominated candidates to keep the search tractable over $\mathcal{P}$. Specifically, a dominance ratio $\lambda \in (0,1]$ is enforced. If the self-play utility of a configuration falls below $\lambda$ times the best observed utility achieved against it under the grid-relevant metrics encoded in $\phi_k$, the configuration is eliminated.

\begin{algorithm}[H]
\caption{Protocol Candidate Screening for Grid-Adaptive NE}
\begin{algorithmic}[1]
\REQUIRE $\mathcal{P}$: protocol parameter set, $\hat{U}_k(\cdot,\cdot)$: SCMC-based grid utility, $\lambda$: screening ratio
\STATE For all $\vartheta \in \mathcal{P}$, estimate $\hat{U}_k(\vartheta, \vartheta)$ via $M_{base}$ SCMC runs.
\STATE Initialize $\mathcal{S}_{remain} \leftarrow \mathcal{P}$; $\mathcal{S}_{final} \leftarrow \emptyset$
\WHILE{$|\mathcal{S}_{remain}| > 1$}
    \STATE Select unexplored $(\vartheta', \vartheta)$ in $\mathcal{P} \times \mathcal{S}_{remain}$
    \STATE Estimate $\hat{U}_k(\vartheta', \vartheta)$ by $M_{step}$ SCMC runs.
    \IF{$\hat{U}_k(\vartheta, \vartheta) < \lambda \cdot \hat{U}_k(\vartheta', \vartheta)$}
        \STATE $\mathcal{S}_{remain} \gets \mathcal{S}_{remain} \setminus \{\vartheta\}$
    \ELSIF{All $\hat{U}_k(\cdot,\vartheta)$ evaluated for $\cdot \in \mathcal{P}$}
        \STATE $\mathcal{S}_{remain} \gets \mathcal{S}_{remain} \setminus \{\vartheta\}$; $\mathcal{S}_{final} \gets \mathcal{S}_{final} \cup \{\vartheta\}$
    \ENDIF
\ENDWHILE
\STATE \textbf{return} $\arg\max_{\vartheta \in \mathcal{S}_{final}} \min_{\vartheta' \in \mathcal{P}} \left( \frac{\hat{U}_k(\vartheta, \vartheta)}{\hat{U}_k(\vartheta', \vartheta)} \right)$
\end{algorithmic}
\end{algorithm}
\vspace{-3mm}

Here, every $\hat{U}_k(\cdot,\cdot)$ is measured by the fraction of SCMC runs where the grid-domain CBTL predicate (including alarm responsiveness, reporting window, and cumulative energy) holds for node $k$.

\subsection{Stage 2: Grid-Constrained NE Statistical Certification}

In the second stage, the objective is not exhaustive equilibrium computation, but rather the statistical certification of whether a selected candidate protocol remains robust (in the sense of $\Delta$-relaxed NE) against selfish deviations, under the operational environment and grid constraints.

Upon identifying a protocol $\vartheta^\star$, we perform statistical validation of $\vartheta^\star$'s robustness as a $\Delta$-relaxed NE under grid constraints. For every alternative $\vartheta' \in \mathcal{P}$,
\begin{equation}
\hat{U}_k(\vartheta^\star, ..., \vartheta^\star) \geq \Delta \cdot \hat{U}_k(\vartheta', \vartheta^\star_{-k})
\end{equation}
where each utility is empirically determined by $M_{cert}$ SCMC runs with realistic grid-side event schedules, node heterogeneity, and parameter uncertainty.

\begin{table*}[!t]
\centering
\caption{%
Performance of IEEE 802.15.4 CSMA/CA Protocol Parameterizations in Power Grid IoT Sensor Networks ($N=20$).
P.S.: Parameter Set; $[BE_{\min}, BE_{\max}]$: Backoff exponent range; $N_{\text{retx}}$: Max retransmissions; CCA: Clear Channel Assessment duration ($\mu$s); D.C.: Duty Cycle (\%); $\hat{U}$: Utility (fraction of successful runs); Delivery Ratio (\%): Status frames delivered within deadlines; Avg. Energy (\ensuremath{\mathrm{mJ}}): Mean per-node energy per cycle; $\Delta_{\text{grid}}$: Robustness coefficient under $\Delta$-relaxed NE.
NE: Nash equilibrium under grid constraints; Opt: maximal utility under symmetric operation.
}
\label{tab:ne-detailed-results}

\resizebox{\textwidth}{!}{%
\begin{tabular}{@{}lcccccccccc@{}}
\toprule
\textbf{P.S.} & $[BE_{\min}, BE_{\max}]$ & $N_{\text{retx}}$ & CCA ($\mu$s) & D.C. (\%) & \textbf{Strategy} & $\hat{U}$ & Avg. Energy (\ensuremath{\mathrm{mJ}}) & Delivery Ratio (\%) & $\Delta_{\text{grid}}$ \\
\midrule
A & [2,4] & 2 & 128 & 8 & NE & 0.911 $\pm$ 0.003 & 147.3 $\pm$ 2.1 & 92.7 $\pm$ 0.6 & 0.970 \\
B & [2,5] & 2 & 128 & 9 & NE & 0.914 $\pm$ 0.002 & 149.2 $\pm$ 2.0 & 93.2 $\pm$ 0.5 & 0.971 \\
C & [3,5] & 3 & 160 & 10 & Opt & 0.943 $\pm$ 0.002 & 149.7 $\pm$ 1.8 & 96.0 $\pm$ 0.4 & 0.973 \\
D & [4,6] & 4 & 192 & 12 & Opt & 0.949 $\pm$ 0.002 & 148.6 $\pm$ 1.7 & 97.2 $\pm$ 0.4 & 0.974 \\
E & [2,3] & 1 & 96 & 7 & -- & 0.862 $\pm$ 0.004 & 142.1 $\pm$ 2.6 & 89.5 $\pm$ 0.8 & -- \\
F & [2,6] & 4 & 192 & 12 & -- & 0.924 $\pm$ 0.003 & 152.8 $\pm$ 2.2 & 92.1 $\pm$ 0.7 & -- \\
\bottomrule
\end{tabular}%
}
\vspace{-4mm}
\end{table*}

Statistical guarantees are established via one-sided confidence bounds for Bernoulli satisfaction outcomes. For each protocol pair, let $\hat{U}_k$ be the empirical success proportion over $M_{\text{cert}}$ independent executions. We compute a lower confidence bound $\underline{U}_k$ for $\hat{U}_k(\vartheta^\star,\vartheta^\star)$ and an upper confidence bound $\overline{U}_k$ for $\hat{U}_k(\vartheta',\vartheta^\star)$ using an exact Clopper-Pearson interval \cite{clopper1934confidence}. To control family-wise error across deviations, we apply a Bonferroni correction with per-comparison level $\alpha_{\text{grid}}/|\mathcal{P}|$.
The certified robustness coefficient is defined as
\begin{equation}
\Delta_{\text{grid}} =
\min_{\vartheta' \in \mathcal{P}}
\frac{\underline{U}_k(\vartheta^\star,\vartheta^\star)}
{\overline{U}_k(\vartheta',\vartheta^\star)}.
\end{equation}
A protocol $\vartheta^\star$ is reported as grid-robust if $\Delta_{\text{grid}}$ exceeds a target threshold and all operational constraints embedded in the CBTL predicates are satisfied under the workload model.

\subsection{Scenario-Specific Implementation}

All SCMC executions are parameterized using grid-derived schedules (maintenance windows, alarm frequency), device-specific energy models, and event injection based on actual or synthetic power disturbance traces. The candidate protocol set $\mathcal{P}$ includes only those configurations compliant with both IEEE 802.15.4 protocol rules and local grid operation guidelines (maximum status frame intervals, minimum online node ratios). CBTL predicates are explicitly engineered to measure metrics prioritized in grid automation, such as latency to alarm broadcast or resilience during repeated blackout cycles.

\vspace{-2mm}
\section{Experimental Evaluation}

This section empirically evaluates the proposed statistical screening and certification framework for IEEE~802.15.4 CSMA/CA parameter selection in power grid IoT sensor networks.

\begin{figure}[!t]
    \centering
    \includegraphics[width=\linewidth]{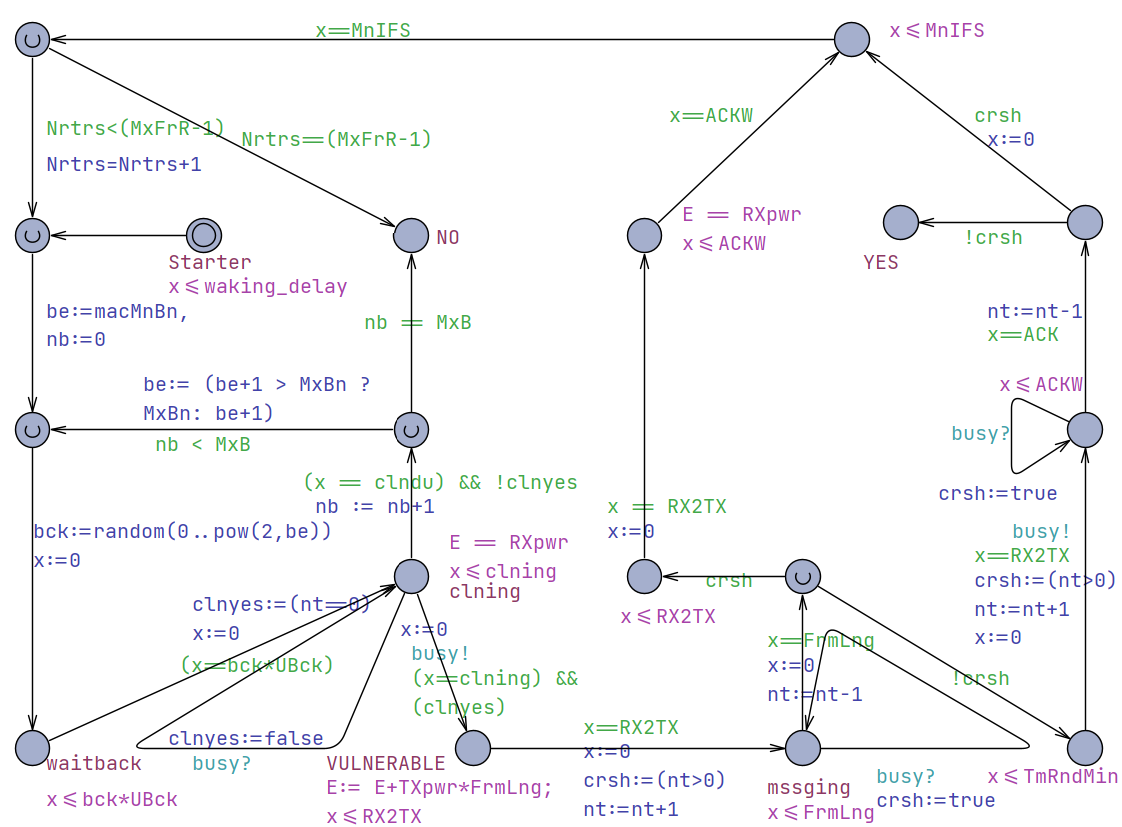}
    \vspace{-4mm}
    \caption{UPPAAL timed automata model of a sensor node implementing the IEEE 802.15.4 CSMA/CA protocol, parameterized for power grid IoT scenarios. The model captures the full protocol state space, including channel contention, backoff, retransmission, and per-state energy accounting, and serves as the core simulation engine for all experimental evaluation and statistical screening in this study.}
    \label{fig:uppaal_model}
    \vspace{-2mm}
\end{figure}

\subsection{Simulation Platform and Computational Resources}

All simulations were conducted using an in-house implementation of the STHA modeling framework, with protocol verification and utility estimation performed via UPPAAL SMC version 5.0 \cite{behrmann2006uppaal,david2015uppaalsmc}. The entire screening and certification process was automated using a Python 3.9 orchestration pipeline. Simulations were executed on a compute cluster equipped with dual Intel Xeon Gold 6338 CPUs and 256 GB RAM, running Ubuntu Server 22.04 LTS.

For each protocol configuration, $M_{\text{screen}} = 10{,}000$ simulation runs were conducted during the screening phase, and $M_{\text{cert}} = 100{,}000$ runs were performed for statistical certification. The total CPU time for the complete evaluation was approximately 950 core-hours. All random seeds, configuration files, and scripts used to reproduce the reported results will be released in an anonymized repository upon acceptance.

\begin{table*}[!t]
\centering
\caption{Baseline comparison of CSMA/CA parameterizations under the same grid workload ($N=20$). Type labels are BL (heuristic baseline), NE (certified Nash equilibrium under grid constraints), and OPT (utility-maximizing symmetric setting). Utility and delivery ratio are reported as mean $\pm$ standard error. ``Budget OK'' indicates whether Avg. Energy is within the 150~\ensuremath{\mathrm{mJ}} per-cycle budget.}
\label{tab:baselines}
\footnotesize
\setlength{\tabcolsep}{3.6pt}
\renewcommand{\arraystretch}{1.10}

\begin{tabular}{@{}lc cccc ccccc@{}}
\toprule
\textbf{Type} & \textbf{Cfg.} &
\multicolumn{4}{c}{\textbf{CSMA/CA parameters}} &
\multicolumn{5}{c}{\textbf{Outcomes under grid workload}} \\
\cmidrule(lr){3-6}\cmidrule(lr){7-11}
& &
$[BE_{\min}, BE_{\max}]$ & $N_{\text{retx}}$ & CCA ($\mu$s) & D.C. (\%) &
$\hat{U}$ & Delivery (\%) & Avg. Energy (mJ) & Budget OK & $\Delta_{\text{grid}}$ \\
\midrule
BL  & E & [2,3] & 1 & 96  & 7  & 0.862 $\pm$ 0.004 & 89.5 $\pm$ 0.8 & 142.1 $\pm$ 2.6 & Yes & n/a \\
BL  & F & [2,6] & 4 & 192 & 12 & 0.924 $\pm$ 0.003 & 92.1 $\pm$ 0.7 & 152.8 $\pm$ 2.2 & No  & n/a \\
\addlinespace[0.4em]
NE  & \textbf{A} & [2,4] & 2 & 128 & 8  & 0.911 $\pm$ 0.003 & 92.7 $\pm$ 0.6 & 147.3 $\pm$ 2.1 & Yes & 0.970 \\
NE  & \textbf{B} & [2,5] & 2 & 128 & 9  & 0.914 $\pm$ 0.002 & 93.2 $\pm$ 0.5 & 149.2 $\pm$ 2.0 & Yes & 0.971 \\
\addlinespace[0.4em]
OPT & \textbf{C} & [3,5] & 3 & 160 & 10 & 0.943 $\pm$ 0.002 & 96.0 $\pm$ 0.4 & 149.7 $\pm$ 1.8 & Yes & 0.973 \\
OPT & \textbf{D} & [4,6] & 4 & 192 & 12 & 0.949 $\pm$ 0.002 & 97.2 $\pm$ 0.4 & 148.6 $\pm$ 1.7 & Yes & 0.974 \\
\bottomrule
\end{tabular}
\end{table*}

\subsection{Experimental Design and Parameterization}

To evaluate the effectiveness of the proposed statistical screening and certification methodology, we conducted a set of systematic experiments focusing on the IEEE 802.15.4 CSMA/CA protocol, under a realistic power grid IoT sensor network scenario. The simulation environment was implemented using the STHA framework, with all parameters reflecting grid operational and regulatory requirements. The underlying protocol dynamics were captured using a UPPAAL-based timed automata model \cite{zhou2025comprehensive}, as shown in Fig.~\ref{fig:uppaal_model}.

The network topology comprises $N=20$ sensor nodes representing intelligent terminal devices deployed within a power substation. Each node is modeled as an independent STHA instance, and nodes communicate over a shared wireless medium. To reflect realistic operational requirements, each node is required to transmit at least one status frame within a reporting interval of $T_{\text{rep}}=45$ seconds. In addition, when abnormal grid events occur, nodes must deliver alarm messages with an end-to-end delay not exceeding $T_{\text{alarm}}=3$ seconds. Energy consumption is constrained by an average budget of $E_{\text{max}}=150$~mJ per reporting cycle, consistent with battery and operational lifecycle limits. Furthermore, to ensure monitoring coverage during blackout or recovery events, at least $90\%$ of sensor nodes are required to remain online, enforcing a minimum availability threshold of $\rho_{\min}=0.9$.

All admissible protocol configurations $\mathcal{P}$ were generated by varying CSMA/CA parameters including backoff exponents $[BE_{\min}, BE_{\max}] \in \{[2,4], [3,5], [4,6]\}$, CCA detection durations $\in \{128, 160, 192\}$~$\mu$s, maximum retransmissions $N_{\text{retx}} \in \{2,3,4\}$, and duty-cycle ratios in the range $8\%$ to $12\%$. Each configuration was verified to comply with both the IEEE 802.15.4 protocol and local grid deployment standards.

\subsection{Model calibration and fidelity checks}

The timed-automata model instantiates IEEE~802.15.4 CSMA/CA primitives using standard-defined timing units and protocol transitions, including backoff countdown, CCA, transmission, ACK waiting, and bounded retransmissions. Energy accounting is implemented as a priced extension in which each protocol location is associated with a mode-dependent power draw and the accumulated cost integrates power over time within each reporting cycle \cite{larsen2005multipriced}. The parameterization follows a direct mapping from protocol settings to model guards and invariants, such that $[BE_{\min},BE_{\max}]$ determines the contention window evolution, $N_{\text{retx}}$ bounds retransmission loops, and the duty-cycle schedule controls sleep and wake transitions. We performed sanity checks to verify that increasing contention windows reduces collision-dominated failures under high load, increasing retransmission limits improves delivery at the cost of higher energy, and more aggressive duty cycling lowers energy while degrading alarm latency.

\subsection{Screening and Certification Methodology}

The experimental workflow consists of two stages. \textit{Stage~1 (Screening):} For each candidate configuration $\vartheta\!\in\!\mathcal{P}$, we estimate the empirical utility $\hat{U}_k(\vartheta,\vartheta)$ as the fraction of $M_{\text{screen}}=10^4$ STHA--SCMC runs in which node $k$ delivers at least one status or alarm message within the mandated $T_{\text{rep}}$ or $T_{\text{alarm}}$ window, does not exceed the cycle energy budget $E_{\text{max}}$, and remains available for at least $90\%$ of blackout-recovery intervals. Configurations violating any operational constraint are discarded, and the remaining candidates are further pruned using a dominance-ratio threshold $\lambda=0.85$. \textit{Stage 2 (Certification):} For each surviving candidate, we applied the grid-constrained $\Delta_{\text{grid}}$-relaxed NE certification procedure described in Section IV, using $M_{\text{cert}}=10^5$ simulation runs per scenario and significance level $\alpha_{\text{grid}}=0.01$. All certification experiments included randomized traffic surges, event-triggered alarms, and heterogeneous node initializations.

\subsection{Uncertainty Quantification}

We report standard errors (SE) for all metrics computed from the same SCMC runs. For probability-type measures (utility, delivery ratio), the SE is $SE=\sqrt{\hat{p}(1-\hat{p})/M}$, where $\hat{p}$ is the empirical proportion and $M$ the number of independent executions. For mean-valued metrics (average energy), the SE is $SE=\hat{s}/\sqrt{M}$ with $\hat{s}$ the sample standard deviation. Unless stated otherwise, two-sided $(1-\alpha)$ confidence intervals are reported as $\hat{\mu}\pm z_{1-\alpha/2}\,SE$.

\subsection{Baseline comparison}
We include two heuristic baselines that reflect common tuning practices under reliability and energy constraints \cite{nagai2024improve,kaddi2024energy}. Baseline E adopts an aggressive backoff with a minimal retry limit, which reduces channel access delay but increases collision risk under contention. Baseline F increases the retry allowance to favor delivery success, at the cost of higher energy consumption and longer active time. As reported in Table~\ref{tab:baselines}, the certified NE configuration \textbf{B} increases utility from 0.862 to 0.914 relative to baseline E, corresponding to a 6.0\% relative gain, and raises the delivery ratio from 89.5\% to 93.2\%. Compared with baseline F, configuration \textbf{B} reduces the mean per-cycle energy from 152.8~mJ to 149.2~mJ, a 2.4\% relative reduction, while maintaining comparable delivery performance. The utility-maximizing symmetric configuration \textbf{D} achieves the highest utility of 0.949, but it requires a larger contention window and a higher duty cycle, which increases the protocol aggressiveness in channel access and activity scheduling.

\section{Conclusion and Discussion}

The proposed framework statistically certifies CSMA/CA parameterizations for power grid sensor networks, ensuring robustness against selfish adaptation while satisfying reliability, timeliness, and energy constraints. The results confirm its practical suitability for critical infrastructure deployment. Future work should improve fidelity by considering multi-channel interference, heterogeneous devices, spatial topology, and resilience to faults and security threats.


\bibliographystyle{IEEEtran}
\bibliography{main}

\end{document}